\documentclass{jpsj2}

\newcommand{\bm}{\boldsymbol{m}}

\newcommand{\bQ}{\boldsymbol{Q}}

\title{Analytic Property of Vortex-Type Spin Structures in Itinerant Electron System with Inversion Symmetry}

\author{Yoshiro Kakehashi\thanks{yok@sci.u-ryukyu.ac.jp, to be published in J. Phys. Soc. Jpn.}}
\inst{Department of Physics, Faculty of Science,\\
University of the Ryukyus, \\
1 Senbaru, Nishihara, Okinawa, 903-0213, Japan} 

\abst{
Analytic properties of the multiple spin density waves showing vortex structures, which have recently been found in the fcc Hubbard model using the generalized Hartree-Fock (GHF) approximation, have been investigated.  It is shown that the 2$Q$ multiple helical spin density waves (SDW) consist of two types of half-skyrmion vortices and two types of antivortices.  In the 3$Q$ multiple helical SDW, the half-skyrmion vortex structures with the ``westerlies'' are shown to be twisted along the axis perpendicular to the vortex plane.  The 12$Q$ multiple SDW (12QMSDW) are analyzed as a superposition of the 2-4QMSDW on the (001) plane and the remaining 4QMSDW.  It is shown analytically that the 2-4QMSDW forms the planar vortices and antivortices with zero core polarization, so that the 12QMSDW forms the same types of vortices with additional SDW polarization in the [001] direction.  These results are in good agreement with the  visualization image data based on the GHF calculations.
}

\kword{ itinerant magnetism, multiple helical spin density waves, vortex spin structure,  magnetic skyrmions, half-skyrmion, antivortex, Hubbard model,
generalized Hartree-Fock approximation }

\begin{document}
\maketitle

\section{Introduction}

The B20-type transition metal compounds such as MnSi~\cite{muhl09} and FeGe~\cite{yu11} are wellknown to show a vortex-type structure called the magnetic skyrmions~\cite{skyr61,skyr62,bogd89,bogd94,bogd01,ross06} in which the magnetic moments being perpendicular to the vortex plane at the core center incline spirally away from the center and point in the opposite direction on the periphery.
Their magnetic structures and properties have been much investigated in the past decade~\cite{fert17,ever18}, since the magnetic skyrmions are a new type of topologically protected system and are expected to be applicable to the nano-size magnetic devices.
More recently, new kind of vortex structure called the half skyrmion, in which the inclined magnetic moments are in the vortex plane on the periphery, has been found in the $\beta$-Mn type Co${}_{8}$Zn${}_{9}$Mn${}_{3}$ compound~\cite{yu18}. In the MnGe compound~\cite{zhan16}, the 3 dimensional type of skyrmions have been discovered.
Moreover a new kind of 4$Q$ multiple spin density waves (MSDW) has been found in MnSi${}_{1-x}$Ge${}_{x}$ ($0.25 < x < 0.7$) alloys~\cite{fuji19}.
These experimental results indicate that a variety of the vortex-type magnetic structures and related MSDW are possible in itinerant electron systems.

Although the Dzyaloshinski-Moriya (DM) interactions~\cite{dzya58,mori60} play an important role in the formation of skyrmions mentioned above~\cite{bak80,yi09,yu10}, recent theoretical investigations based on the Heisenberg model have shown that the skyrmion-type lattice structures can also be stabilized by the competition between the ferro- and antiferro-magnetic interactions~\cite{okubo12}.  
The skyrmion type structure without the DM interactions should be stabilized also in itinerant electron system because more complex competitions between the long-range ferro- and antiferro-magnetic interactions are expected via electron hopping process.

We recently investigated  the real-space magnetic structures of the multiple spin density waves (MSDW) on the fcc lattice using the application visualization system (AVS) image technique to understand the skyrmion-type magnetic structure in itinerant electron system~\cite{kake18}, and found that the skyrmion-type vortex structures are described by the multiple helical SDW.  Using the phenomenological Ginzburg-Landau theory~\cite{uchida06,uchida07,kake13}, we showed that the skyrmion type vortex structures are possible even in the itinerant electron system with inversion symmetry.

In the subsequent paper~\cite{kake20}, which we refer to I hereafter, we performed the self-consistent magnetic-structure calculations for the Hubbard model~\cite{hub63,hub641,hub642,gutz63,gutz64,gutz65,kana63} on the fcc lattice using the Generalized Hartree-Fock (GHF) approximation to investigate the possibility of the itinerant-electron skyrmions from a microscopic point of view.  Assuming various multiple-spin-density waves structures, we examined their stability in the space of the Coulomb interaction energy $U$ and the electron number per site $n_{e}$.  We found the 2$Q$ multiple helical SDW (2QH), the 3$Q$ multiple helical SDW (3QH), and the 12$Q$ multiple SDW (12QMSDW) showing the vortex-type skyrmion structures.  

Although we have shown that the vortex-type skyrmion structures are possible even in the itinerant electron system with inversion symmetry, it is based on the observation of the magnetic structures with use of the AVS image technique.  Therefore, the details of their vortex structures as well as the type of the skyrmion structures have not yet been well understood.  We need alternative approach to clarify the magnetic structures obtained from the numerical calculations.

In this paper, we examine the analytic property of the vortex structures obtained by the self-consistent GHF calculations for the Hubbard model on the fcc lattice.  We start from the principal MSDW obtained by the Fourier analysis of the calculated magnetic structures, and determine the vortex core positions assuming the vortex magnetic structure.  Expanding the MSDW with respect to the position vector around the vortex cores, we determine analytically the local structure of the vortices.

We determined the local magnetic structures of the 2QH, the 3QH, and the 12QMSDW obtained in the self-consistent GHF calculations.  We found that the 2QH consists of the two types of vortices and antivortices.  The former vortices are the half skyrmions~\cite{merm79,naga13}, and form the antiferromagnetic (AF) lattice. The 3QH forms the half-skyrmion-base vortex  structures being twisted along the axis perpendicular to the vortex plane.  The 12QMSDW also form the vortex lattice structure, but these vortices have no core polarization and are not the skyrmion system.  Moreover we found that the two types of antivortices with no core polarization appear in common to the 2QH and the 12QMSDW, while the core centers in the 3QH have the magnetic moments which helically rotate along the axis perpendicular to the vortex plane.

In the following section,  we summarize the magnetic-structure diagram on the fcc lattice obtained in the self-consistent GHF calculations~\cite{kake20}.  New data points are added to the diagram, and a new type of the AF-base half-skyrmion structure is reported there.  In Sect. 3.1, the local structures of the 2QH  is analyzed, and the boundaries of the half-skyrmions are analytically determined.  
In Sect. 3.2, the 3QH structure is analyzed on the basis of the 2QH local structures.  Analytic expressions of the half skyrmions with the ``westerlies'' are given.
Section 3.3 is devoted to the analysis of the 12QMSDW.  We regard the 12QMSDW as a superposition of  the 2-4QMSDW (or the 8QMSDW) and the remaining 4QMSDW, and analyze first the former, taking the same steps as in the 2QH analysis.  We will show that the 2-4QMSDW consists of the vortex and antivortex structures with zero core polarization, so that the 12QMSDW forms the vortex structures which are different from the half skyrmions.
In the last section 4, we summarize the present work and discuss the remaining problems. 
%
%
\begin{figure}[htbp]
\begin{center}
\includegraphics[width=13cm]{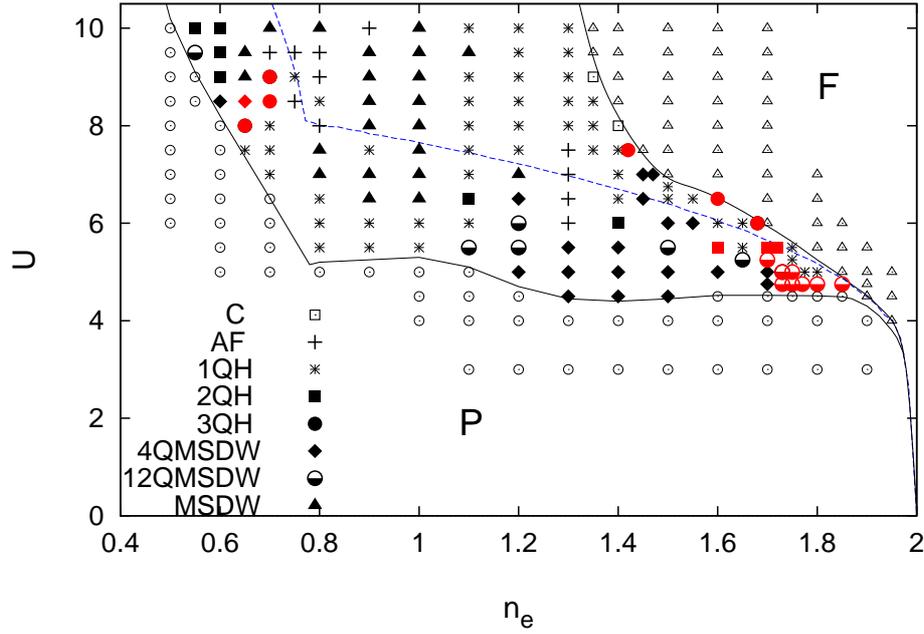}
\end{center}
\vspace{0cm}
\caption{(Color online) Magnetic structure diagram on the $U$-$n_{e}$ plane obtained by the self-consistent generalized Hartree-Fock (GHF) approximation.  F: ferromagnetic state (open triangles), P: paramagnetic state (open circles), C: conical state, AF: anti-ferromagnetic state,  1QH: 1$Q$ helical state, 2QH: 2$Q$ multiple helical spin density wave  (2$Q$-MHSDW) state, 3QH: 3$Q$-MHSDW state, 4QMSDW: 4$Q$ multiple spin density wave (MSDW) state, 12QMSDW: 12$Q$-MSDW state, and MSDW: the other-type MSDW states.  The Coulomb interaction energy parameter $U$ is measured in unit of the nearest-neighbor transfer integral $|t|$.  The vortex-type structures with long wave length are shown by red color.  The paramagnetic and ferromagnetic boundaries are shown by thin curves. The Stoner instability line is shown by the blue dashed line. See paper I (Ref. 24) for more details of magnetic structures. 
}
\label{fg1unpd}
\end{figure}
%
%

\section{Magnetic-Structure Diagram}

We considered in paper I~\cite{kake20} the Hubbard model~\cite{hub63,hub641,hub642,gutz63,gutz64,gutz65,kana63}  on the fcc lattice with nearest neighbor transfer integral $t$ and Coulomb interaction energy $U$, and performed the magnetic structure calculations as a function of $U$ and $n_{e}$ on the basis of the GHF approximation.  We adopted a large cluster consisting of the $10 \times 10 \times 10$ fcc unit cells with periodic boundary condition and applied the recursion method~\cite{hay75,heine80} to calculate the electronic states of magnetic structures.
  
We considered, as the input magnetic structures for self-consistent calculations, the single-$Q$ helical state (1QH), the 2QH, the 3QH, as well as the paramagnetic state (P) and the ferromagnetic state (F).  Starting from these structures and varying the wave numbers, we performed the self-consistent GHF calculations at zero temperature and determined the stable magnetic structure comparing their total energies. 
Since the output magnetic structures do not necessarily agree with the starting ones, we also obtained the other magnetic structures in these calculations.

We summarize the magnetic structures obtained in paper I in Fig. 1.  Note that we increased the number of points in the figure, making further numerical calculations, in order
 to make clearer the stability regions in the diagram.
In the upper-right and the lower-left  regions, we find the F and P states, respectively.
Between the F and P, the conical structure (C), the antiferromagnetic structure of the first and second kinds (AF), the 1QH, the 2QH, the 4$Q$ multiple spin density waves (4QMSDW), the 12$Q$ multiple SDW (12QMSDW), and the other MSDW are stabilized due to competition between the ferro- and antiferro-magnetic interactions via electron hopping.
The wave numbers of these SDW and MSDW show the maximum around $n_{e}=1$, and decrease with increasing $n_{e}$, since the ferromagnetic interactions become dominant when $n_{e}$ approaches the ferromagnetic boundary.
As we have reported in I, we observed with use of the AVS image analysis, the vortex-type structures with long wave length in the 2QH, 3QH, and 12QMSDW, which are indicated by the  red color in Fig. 1.
These  structures appear near the Stoner ferromagnetic boundary.
 
 The 3QH around $(n_{e},U) = (0.7, 8.5)$ are the AF-base vortex structure.  We added a new point of the 4QMSDW with red color at $(n_{e}, U) = (0.65, 8.5)$, because it shows the AF-base vortex structure of the 2QH type.  In fact, according to the Fourier analysis of the structure, the magnetic moment on site $\boldsymbol{R}_{l} = (x_{l}, y_{l}, z_{l})$ is given by a 4QMSDW as follows.
\begin{align}
\bm_{l} &= m_{1} \boldsymbol{j} \cos \tilde{\boldsymbol{Q}}_{1} \! \cdot \! \boldsymbol{R}_{l} +
m_{2}\boldsymbol{k} \sin \tilde{\boldsymbol{Q}}_{1} \! \cdot \! \boldsymbol{R}_{l} +
m_{2}\boldsymbol{k} \cos \tilde{\boldsymbol{Q}}_{2} \! \cdot \! \boldsymbol{R}_{l} +
m_{1}\boldsymbol{i} \sin \tilde{\boldsymbol{Q}}_{2} \! \cdot \! \boldsymbol{R}_{l}  \nonumber \\
&+ m_{1} \boldsymbol{j} \cos \tilde{\boldsymbol{Q}}_{3} \! \cdot \! \boldsymbol{R}_{l} -
m_{2}\boldsymbol{k} \sin \tilde{\boldsymbol{Q}}_{3} \! \cdot \! \boldsymbol{R}_{l} +
m_{2}\boldsymbol{k} \cos \tilde{\boldsymbol{Q}}_{4} \! \cdot \! \boldsymbol{R}_{l} -
m_{1}\boldsymbol{i} \sin \tilde{\boldsymbol{Q}}_{4} \! \cdot \! \boldsymbol{R}_{l}.
\label{af4qm}
\end{align}
Here $m_{1}=0.131$ and $m_{2}=0.068$.  $\boldsymbol{i}$, $\boldsymbol{j}$, $\boldsymbol{k}$ are the unit vectors for the $x$, $y$, and $z$ axes, respectively.   The wave vectors $\{ \tilde{\boldsymbol{Q}}_{n} \}$ are given by $\tilde{\boldsymbol{Q}}_{1}=(q, 0, 1)$, $\tilde{\boldsymbol{Q}}_{2}=(0, q, 1)$, $\tilde{\boldsymbol{Q}}_{3}=(-q, 0, 1)$, $\tilde{\boldsymbol{Q}}_{4}=(0, -q, 1)$ in unit of $2\pi/a$, and the wave number $q=0.3$, $a$ being the lattice constant.   
The magnetic moments (\ref{af4qm}) on the $n$th layer ( , {\it i.e.}, $z_{l}=na/2$) can be folded as follows.
\begin{align}
\bm_{l} &= (-)^{n} 2 \, (m_{1} \, \boldsymbol{j} \cos \boldsymbol{Q}_{1} \! \cdot \! \boldsymbol{R}_{l} +
m_{2} \, \boldsymbol{k} \sin \boldsymbol{Q}_{1} \! \cdot \! \boldsymbol{R}_{l}  \nonumber \\
&\hspace{13mm}  + m_{2} \, \boldsymbol{k} \cos \boldsymbol{Q}_{2} \! \cdot \! \boldsymbol{R}_{l} +
m_{1} \, \boldsymbol{i} \sin \boldsymbol{Q}_{2} \! \cdot \! \boldsymbol{R}_{l} ) \, .
\label{af2qm}
\end{align}
Here $\boldsymbol{Q}_{1}=(q, 0, 0)$, $\boldsymbol{Q}_{2}=(0, q, 0)$.
This is an AF-base 2QH half-skyrmion structure~\cite{kake18,kake20}, in which the magnetic moments on the (001) plane change their sign alternatively along the $z$ axis. 
The AF-base 2QH is a frustrated system which does not show any dip at the Fermi level in the density of states~\cite{kake20}.

In the following sections, we clarify the vortex structures of the 2QH around $(n_{e}, U)=(1.70-1.73, 5.5)$, the 3QH around $(n_{e}, U)=(1.68, 6)$, $(1.6, 6.5)$, and $(1.42, 7.5)$, and  the 12QMSDW around $(n_{e}, U)=(1.73, 5)$.

\section{Analysis of Vortex-Type Structures}

\subsection{2$Q$ multiple helical SDW}

We found in I the half-skyrmion type vortex structure on the $xy$ plane at $(n_{e}, U)=(1.70-1.73, 5.5)$ using the AVS image analysis.  The principal terms of the Fourier analysis for the calculated magnetic moments $\bm_{l}$ are given by two elliptical-helical waves as follows.
\begin{align}
\bm_{l} &= m_{1} \, \boldsymbol{j} \cos \boldsymbol{Q}_{1} \! \cdot \! \boldsymbol{R}_{l} +
m_{2} \, \boldsymbol{k} \sin \boldsymbol{Q}_{1} \! \cdot \! \boldsymbol{R}_{l} \nonumber \\
& \ +
m_{2} \, \boldsymbol{k} \cos \boldsymbol{Q}_{2} \! \cdot \! \boldsymbol{R}_{l} +
m_{1} \, \boldsymbol{i} \sin \boldsymbol{Q}_{2} \! \cdot \! \boldsymbol{R}_{l} \, .
\label{2qm}
\end{align}
Here $m_{1}=0.075$, $m_{2}=0.044$ for $(n_{e}, U)=(1.7, 5.5)$, and the wave vectors are given by $\boldsymbol{Q}_{1}=(q, 0, 0)$, $\boldsymbol{Q}_{2}=(0, q, 0)$, and $q=0.2$. We neglect in the following analysis additional 4$Q$-MSDW satellite terms found in the Fourier analysis in I.

The vortices are accompanied by the vortex cores.  Assuming the existence of the skyrmion-type vortex structure on the $xy$ plane, the condition for the core center $\boldsymbol{R}_{c}$ is given by
\begin{align}
\cos \boldsymbol{Q}_{1} \! \cdot \! \boldsymbol{R}_{c} =
\sin \boldsymbol{Q}_{2} \! \cdot \! \boldsymbol{R}_{c} = 0 \, .
\label{2qhccond}
\end{align}
Solving the above equations, we obtain $\boldsymbol{R}_{c}$ as
\begin{align}
\boldsymbol{R}_{c} = \Big( \frac{(2n+1)L}{4}, \frac{n^{\prime}L}{2}, z_{l} \Big) \, .
\label{2qhrc}
\end{align}
Here $n$ and $n^{\prime}$ are integer, $L$ is the wave length defined by $L=a/q$, and $z_{l}$ is the $z$ coordinate of the lattice points on a $xy$ plane.  We note that $\boldsymbol{R}_{c}$ are not necessarily located on the lattice points in general, though it is not essential in the following analysis.

Introducing the lattice point $\boldsymbol{r}_{l} = (\hat{x}_{l}, \hat{y}_{l}, \hat{z}_{l})$ measured from the core center $\boldsymbol{R}_{c}$ ($\boldsymbol{R}_{l}=\boldsymbol{r}_{l}+\boldsymbol{R}_{c}$), Eq. (\ref{2qm}) is written as
\begin{align}
\bm_{l} &= m_{1} \boldsymbol{j} \cos (\boldsymbol{Q}_{1} \! \cdot \! \boldsymbol{r}_{l} + \boldsymbol{Q}_{1} \! \cdot \! \boldsymbol{R}_{c}) +
m_{2}\boldsymbol{k} \sin (\boldsymbol{Q}_{1} \! \cdot \! \boldsymbol{r}_{l} + \boldsymbol{Q}_{1} \! \cdot \! \boldsymbol{R}_{c}) \nonumber \\
& \ +
m_{2}\boldsymbol{k} \cos (\boldsymbol{Q}_{2} \! \cdot \! \boldsymbol{r}_{l}+\boldsymbol{Q}_{2} \! \cdot \! \boldsymbol{R}_{c}) +
m_{1}\boldsymbol{i} \sin (\boldsymbol{Q}_{2} \! \cdot \! \boldsymbol{r}_{l}+\boldsymbol{Q}_{2} \! \cdot \! \boldsymbol{R}_{c}) \, .
\label{2qm2}
\end{align}
When $n$ and $n^{\prime}$ are even, {\it i.e.}, $n=2k$ and $n^{\prime}=2k^{\prime}$, we have
\begin{align}
\bm_{l} &= -m_{1} \boldsymbol{j} \sin \boldsymbol{Q}_{1} \! \cdot \! \boldsymbol{r}_{l} +
m_{2}\boldsymbol{k} \cos \boldsymbol{Q}_{1} \! \cdot \! \boldsymbol{r}_{l} \nonumber \\
& \ \ \ +
m_{2}\boldsymbol{k} \cos \boldsymbol{Q}_{2} \! \cdot \! \boldsymbol{r}_{l} +
m_{1}\boldsymbol{i} \sin \boldsymbol{Q}_{2} \! \cdot \! \boldsymbol{r}_{l} \, .
\label{2qmee}
\end{align}
Expanding $\bm_{l}$ with respect to $\boldsymbol{Q}_{1} \! \cdot \! \boldsymbol{r}_{l}$ and $\boldsymbol{Q}_{2} \! \cdot \! \boldsymbol{r}_{l}$ up to the second order, we obtain
\begin{align}
\bm_{l} &= m_{1}q (\hat{y}_{l} \boldsymbol{i} - \hat{x}_{l} \boldsymbol{j}) +
2m_{2} \Big[1 - \frac{1}{4}q^{2}(\hat{x}_{l}^{2}+\hat{y}_{l}^{2}) \Big] \boldsymbol{k} \, .
\label{2qmee1}
\end{align}
With use of the polar coordinates $(r_{l}, \phi_{l})$ on the $xy$ plane, we obtain
\begin{align}
\bm_{l} &= m_{1}q r_{l} \Big[ \cos \Big( \phi_{l} - \frac{\pi}{2} \Big) \boldsymbol{i} + \sin \Big( \phi_{l} - \frac{\pi}{2} \Big) \boldsymbol{j} \Big] +
2m_{2} \Big(1 - \frac{1}{4}q^{2} r_{l}^{2} \Big) \boldsymbol{k} \, .
\label{2qmee2}
\end{align}
This is a clockwise vortex with the winding number $w=1$, helicity $\alpha=-\pi/2$, and core polarization $p_{\rm c} = 2m_{2}$ which is perpendicular to the vortex plane.
The vortex structure may be schematically expressed by (I) in Fig. 2.
%
%
\begin{figure}[htbp]
\begin{center}
\includegraphics[width=12cm]{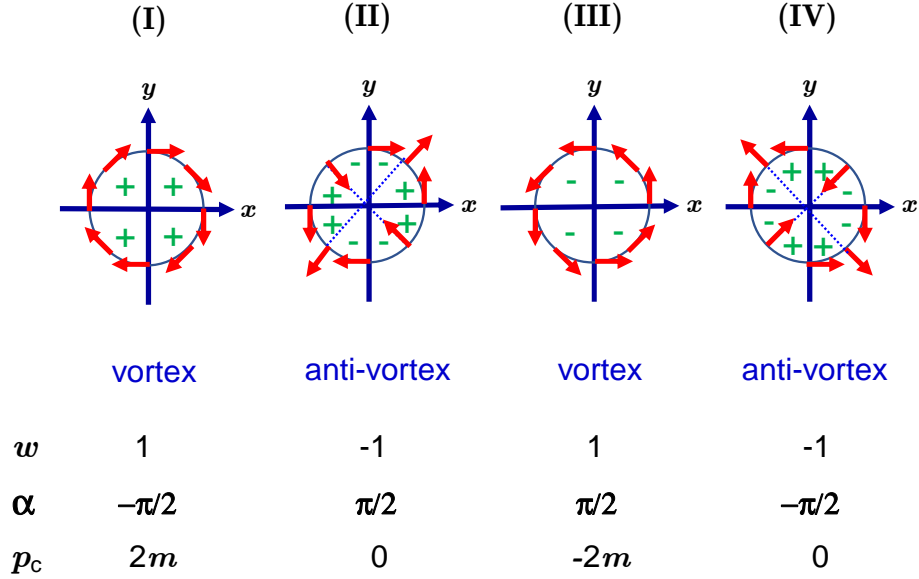}
\end{center}
\vspace{0cm}
\caption{ (Color online)  4-types of vortex structures in the 2QH.
(I): vortex structure with winding number $w=1$, helicity $\alpha=-\pi/2$, and core polarization $p_{\rm c}=2m$ centered at $\boldsymbol{R}_{\rm c}=((k+1/4)L, k^{\prime}L, z_{l})$, (II): anti-vortex structure with $w=-1$, $\alpha=\pi/2$, and $p_{\rm c}=0$ centered at $\boldsymbol{R}_{\rm c}=((k+3/4)L, k^{\prime}L, z_{l})$,  (III): vortex structure with $w=1$, $\alpha=\pi/2$, and $p_{\rm c}=-2m$ centered at $\boldsymbol{R}_{\rm c}=((k+3/4)L, (k^{\prime}+1/2)L, z_{l})$, (IV): anti-vortex structure with $w=-1$, $\alpha=-\pi/2$, and $p_{\rm c}=0$ centered at $\boldsymbol{R}_{\rm c}=((k+1/4)L, (k^{\prime}+1/2)L, z_{l})$. Here $m$ is the amplitude of SDW, $k$, $k^{\prime}$ are integers, $L$ is the wave length of the system ($L=a/q$), and $z_{l}$ is the $z$ coordinate of the $xy$ plane.  The signs inside each circle denote the positive ($+$) or negative ($-$) $z$ polarization being perpendicular to the $xy$ plane.
}
\label{fg2vortype}
\end{figure}
%
%

When $n$ is odd ($n=2k+1$) and $n^{\prime}$ is even ($n^{\prime}=2k^{\prime}$) in $\boldsymbol{R}_{c}$, Eq. (\ref{2qm2}) is expressed as
\begin{align}
\bm_{l} &= m_{1} \boldsymbol{j} \sin \boldsymbol{Q}_{1} \! \cdot \! \boldsymbol{r}_{l} -
m_{2}\boldsymbol{k} \cos \boldsymbol{Q}_{1} \! \cdot \! \boldsymbol{r}_{l} \nonumber \\
& \ +
m_{2}\boldsymbol{k} \cos \boldsymbol{Q}_{2} \! \cdot \! \boldsymbol{r}_{l} +
m_{1}\boldsymbol{i} \sin \boldsymbol{Q}_{2} \! \cdot \! \boldsymbol{r}_{l} \, .
\label{2qmoe}
\end{align}
Taking the same steps as before, we obtain
\begin{align}
\bm_{l} &= m_{1}q r_{l} \Big[ \cos \Big( -\phi_{l} + \frac{\pi}{2} \Big) \boldsymbol{i} + \sin \Big( -\phi_{l} + \frac{\pi}{2} \Big) \boldsymbol{j} \Big] +
\frac{1}{2}m_{2}q^{2}r_{l}^{2} (\cos^{2} \phi_{l} - \sin^{2} \phi_{l} ) \, \boldsymbol{k} \, .
\label{2qmoe2}
\end{align}
This is an antivortex with $w=-1$, $\alpha=\pi/2$, and $p_{\rm c} = 0$.  The antivortex structure may be expressed by (II) in Fig. 2.

Similarly, for $n=2k+1$ and $n^{\prime}=2k^{\prime}+1$, we have
\begin{align}
\bm_{l} &= m_{1}q r_{l} \Big[ \cos \Big( \phi_{l} + \frac{\pi}{2} \Big) \boldsymbol{i} + \sin \Big( \phi_{l} + \frac{\pi}{2} \Big) \boldsymbol{j} \Big] -
2m_{2} \Big(1 - \frac{1}{4}q^{2} r_{l}^{2} \Big) \boldsymbol{k} \, .
\label{2qmoo2}
\end{align}
This is the anticlockwise vortex with $w=1$, $\alpha=\pi/2$, and $p_{\rm c} = -2m_{2}$.  It may be expressed by (III) in Fig. 2.

In the case of $n=2k$ and $n^{\prime}=2k^{\prime}+1$, we obtain
\begin{align}
\bm_{l} &= m_{1}q r_{l} \Big[ \cos \Big( -\phi_{l} - \frac{\pi}{2} \Big) \boldsymbol{i} + \sin \Big( -\phi_{l} - \frac{\pi}{2} \Big) \boldsymbol{j} \Big] -
\frac{1}{2}m_{2}q^{2}r_{l}^{2} (\cos^{2} \phi_{l} - \sin^{2} \phi_{l} ) \, \boldsymbol{k} \, .
\label{2qmeo2}
\end{align}
This is the antivortex with $w=-1$, $\alpha=-\pi/2$, $p_{\rm c} = 0$, and may be expressed by (IV) in Fig. 2.  The 4 types of vortex structures mentioned above are summarized in Fig. 2.

The vortex structures (I) and (III) in Fig. 2 do not form the skyrmions when $r_{l}$ is increased, because Eq. (\ref{2qhccond}) does not have the other solutions with the magnetic moments being antiparallel to the core polarization.  When we assume that Eq. (\ref{2qm}) expresses the half-skyrmion lattice, the boundaries $\boldsymbol{R}=(x, y, z)$ for the half skyrmions are determined by the following condition (, {\it i.e.}, vanishment of the $z$ component of $\bm_{l}$ in Eq. (\ref{2qm}) ):
\begin{align}
\sin \boldsymbol{Q}_{1} \! \cdot \! \boldsymbol{R} +
\cos \boldsymbol{Q}_{2} \! \cdot \! \boldsymbol{R} = 0 \, .
\label{2qhbcond}
\end{align}
This yields two types of linear functions which are orthogonal each other.
\begin{align}
y = x - \big( l - \frac{1}{4} \big) L ,    \\
y = -x - \big( l^{\prime} + \frac{3}{4} \big) L \, .
\label{2qhb}
\end{align}
Here integers $l$ and $l^{\prime}$ are parameters to be determined.
One can determine the boundary for each half skyrmion from these lines.  For example, for the half-skyrmion (III) centered at $\boldsymbol{R}_{c} = ( (k+3/4)L, (k^{\prime}+1/2)L, z_{l} )$, we obtain the boundary lines $y = x - (k-k^{\prime}-1/4) L$, $y = x - (k-k^{\prime}+3/4) L$, $y = -x - (k+k^{\prime}+7/4) L$, and $y = -x - (k+k^{\prime}+3/4) L$.

Figure 3 shows the AVS image on the $xy$ plane  for $(n_{e},U)=(1.7, 5.5)$ and the analytic result for $k=k^{\prime}=1$ and $q=0.2$.  We verify that the analytic results describe the half-skyrmion half-antiskyrmion pairs as well as their boundaries.  In addition, we find the antivortex structures (II) and (IV) between the half-skyrmion particles with the same helicity.  The antivortices were overlooked in the previous AVS-image observation in paper I. 
%
%
\begin{figure}[htbp]
\begin{center}
\includegraphics[width=16cm]{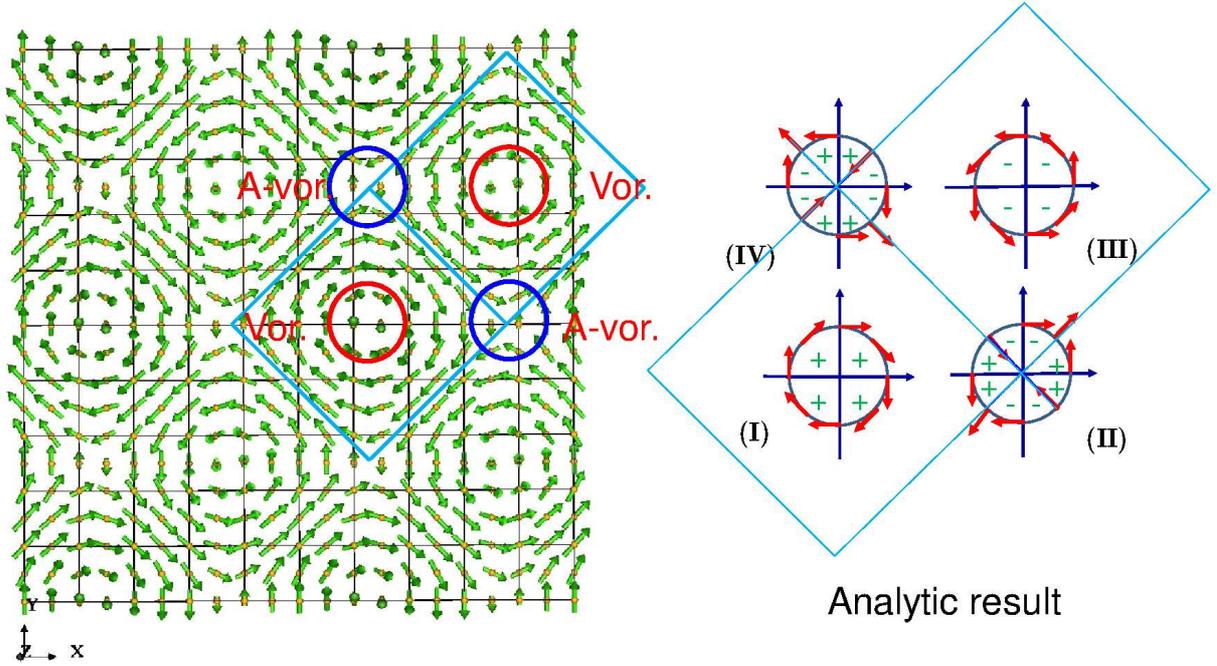}
\end{center}
\vspace{0cm}
\caption{ (Color online)  2$Q$ multiple helical SDW (2QH) on the $xy$ plane obtained by the GHF calculations for $(n_{e},U)=(1.7, 5.5)$ (left) and analytic result (right).  At the left-hand-side, the AVS image on the $10 \times 10 \times 10$ fcc unit cells is shown.  Two types of vortex structures (Vor.) and two types of antivortex structures (A-vor.) are described by the analytic results of (I), (II), (III), and (IV) given in Fig. 2 as shown at the right-hand side.  The blue lines express the half-skyrmion boundaries of (I) and (III).  Use a zoom-in tool to see more detailed structure in the AVS image.
}
\label{fg32qh}
\end{figure}
%
%

\subsection{3$Q$ multiple helical SDW}
The 3$Q$ multiple helical SDW (3QH) with the long wave length have been found at $(n_{e},U) = (1.68, 6)$, $(1.6, 6.5)$, and $(1.42, 7.5)$.  The principal terms of these MSDW are given by
\begin{align}
\bm_{l} &= m \, \big( 
\boldsymbol{j} \cos \boldsymbol{Q}_{1} \! \cdot \! \boldsymbol{R}_{l} +
\boldsymbol{k} \sin \boldsymbol{Q}_{1} \! \cdot \! \boldsymbol{R}_{l} +
\boldsymbol{k} \cos \boldsymbol{Q}_{2} \! \cdot \! \boldsymbol{R}_{l} +
\boldsymbol{i} \sin \boldsymbol{Q}_{2} \! \cdot \! \boldsymbol{R}_{l} \nonumber \\
&\ \ \ \ \ \, +
\boldsymbol{i} \cos \boldsymbol{Q}_{3} \! \cdot \! \boldsymbol{R}_{l} +
\boldsymbol{j} \sin \boldsymbol{Q}_{3} \! \cdot \! \boldsymbol{R}_{l} \big) \ .
\label{3qm}
\end{align}
Here $m=0.067$ for $(n_{e},U) = (1.68, 6)$. The wave vectors $\bQ_{1}$, $\bQ_{2}$, and $\bQ_{3}$ are defined by $\bQ_{1}=(q,0,0)$, $\bQ_{2}=(0,q,0)$, $\bQ_{3}=(0,0,q)$, and $q=0.2$.

The 3QH is regarded as a superposition of the 2QH given by Eq. (\ref{2qm}) with $m_{1}=m_{2}=m$ and the single-$Q$ helical wave $m\boldsymbol{e}(z_{l})$, where $\boldsymbol{e}(z_{l})=\boldsymbol{i} \cos \boldsymbol{Q}_{3} \! \cdot \! \boldsymbol{R}_{l} +
\boldsymbol{j} \sin \boldsymbol{Q}_{3} \! \cdot \! \boldsymbol{R}_{l}$.
Thus the 3QH forms the half-skyrmion vortex structures with a ``westerlies'' $m\boldsymbol{e}(z_{l})$~\cite{kake20}.  The local structure, for example, around $\boldsymbol{R}_{\rm c} = ((k+1/4)L, k^{\prime}L, 0)$ is obtained from Eq. (\ref{2qmee2}) as follows.
\begin{align}
\bm_{l} &= mq r_{l} \Big[ \cos \Big( \phi_{l} - \frac{\pi}{2} \Big) \boldsymbol{i} + \sin \Big( \phi_{l} - \frac{\pi}{2} \Big) \boldsymbol{j} \Big] +
2m \Big(1 - \frac{1}{4}q^{2} r_{l}^{2} \Big) \boldsymbol{k}    \nonumber \\
&+ m \Big(1-\frac{1}{2}q^{2}\hat{z}_{l}^{2} \Big) \boldsymbol{i} + mq\hat{z}_{l} \boldsymbol{j} \, .
\label{3qmee2}
\end{align}

The AVS image obtained for $(n_{e}, U) = (1.68, 6)$ with $q=0.2$ is shown in Fig. 4.  Analytic results of the half-skyrmion vortex structures with the ``westerlies'' $m\boldsymbol{i}$, which are centered at $\boldsymbol{R}_{\rm c} = (25a/4, 5a, 0)$ and $\boldsymbol{R}_{\rm c} = (35a/4, 15a/2, 0)$, respectively, are indicated by the curved arrow circles I and III on the $xy$ plane ($z_{l}=0$).   
The ``westerlies'' $m\boldsymbol{e}(z_{l})$ rotates  on the $xy$ plane with the translation along the $z$ axis.  Therefore the vortices with ``westerlies'', I and III, also rotate along the $z$ axis with the period $L=5a$.  Their magnetic moments at $\boldsymbol{R}_{\rm c}$ would  conically rotate along the $z$ axis; $\bm_{l}=m\boldsymbol{e}(z_{l}) \pm 2m\boldsymbol{k}$.
The antivortices with ``westerlies'', II and IV, obtained by the present analysis, also rotate along the $z$ axis, but the magnetic moments at $\boldsymbol{R}_{\rm c}$ would show the helical rotation along the $z$ axis.
The same local structures appear on the $yz$ and $zx$ planes because Eq. (\ref{3qm}) is invariant for the cyclic transformation of the $x$, $y$, and $z$ coordinates.
%
%
\begin{figure}[htbp]
\begin{center}
\includegraphics[width=17cm]{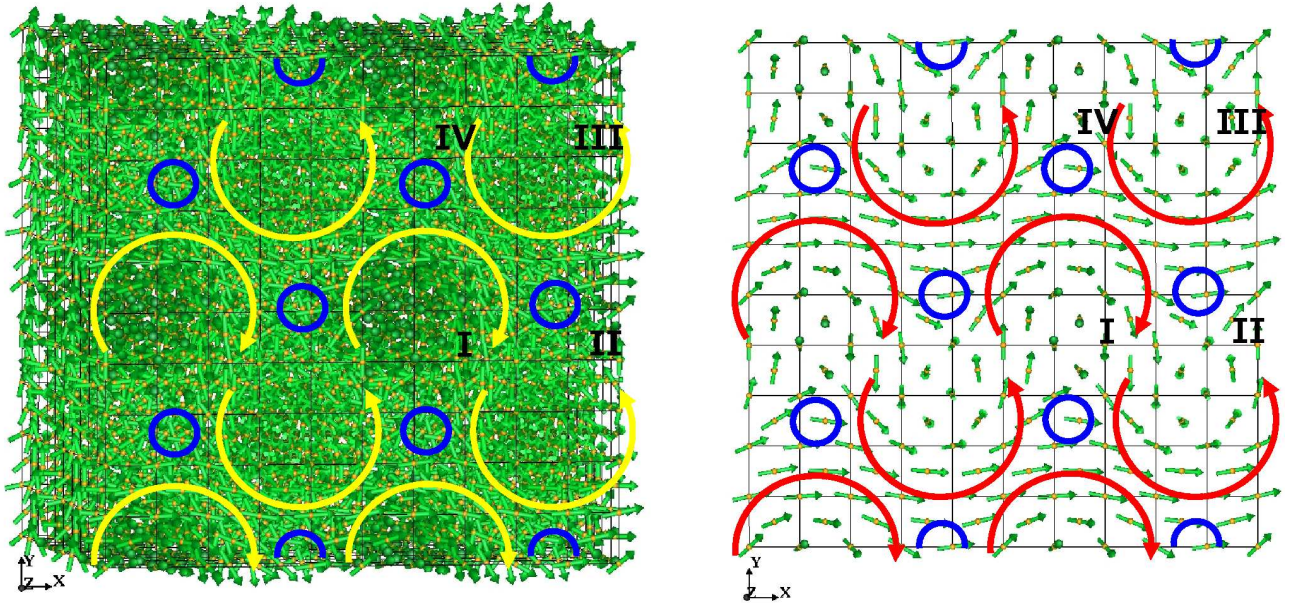}
\end{center}
\vspace{0cm}
\caption{ (Color online)  3$Q$ multiple helical SDW (3QH) on the $xy$ plane obtained by the GHF calculations for $(n_{e},U) = (1.68, 6)$ (left), and its magnetic structure on the $z_{l}=0$ plane (right).  Half-skyrmion vortices with the ``westerlies'' $m\boldsymbol{i}$ (I and III) and antivortices (II and IV) found by the present analysis are shown by the curved arrow circles with yellow (left) or red (right) color, and the blue circles, respectively.  Use a zoom-in tool to see more detailed structure.
}
\label{fg43qh}
\end{figure}
%
%

\subsection{12$Q$ multiple helical SDW}

Alternative MSDW with vortex structures are stabilized around $(n_{e}, U)=(1.75,5)$.  These states are the 12QMSDW~\cite{kake20} given by
\begin{align}
\bm_{l} &= 
m \boldsymbol{i} \big[ -\cos (\bQ_{11} \! \cdot \! \boldsymbol{R}_{l} + \delta) +
\cos (\bQ_{21} \! \cdot \! \boldsymbol{R}_{l}  + \delta) \nonumber \\
&\ \ \ \ \ \ \ \ \ +
\cos (\bQ_{31} \! \cdot \! \boldsymbol{R}_{l}  + \delta) - 
\cos (\bQ_{41} \! \cdot \! \boldsymbol{R}_{l}  + \delta) \big] \nonumber \\
& +
m \boldsymbol{j} \big[ -\cos (\bQ_{12} \! \cdot \! \boldsymbol{R}_{l}  + \delta) +
\cos (\bQ_{22} \! \cdot \! \boldsymbol{R}_{l}  + \delta) \nonumber \\
&\ \ \ \ \ \ \ \ \ +
\cos (\bQ_{32} \! \cdot \! \boldsymbol{R}_{l}  + \delta) - 
\cos (\bQ_{42} \! \cdot \! \boldsymbol{R}_{l}  + \delta) \big] \nonumber \\
& +
m \boldsymbol{k} \big[ -\cos (\bQ_{13} \! \cdot \! \boldsymbol{R}_{l}  + \delta) +
\cos (\bQ_{23} \! \cdot \! \boldsymbol{R}_{l}  + \delta) \nonumber \\
&\ \ \ \ \ \ \ \ \ +
\cos (\bQ_{33} \! \cdot \! \boldsymbol{R}_{l}  + \delta) - 
\cos (\bQ_{43} \! \cdot \! \boldsymbol{R}_{l}  + \delta) \big] \ .
\label{12qsdw}
\end{align}
Here $m=0.022$ for $(n_{e}, U)=(1.75,5)$, $\delta=-\pi/4$, $\bQ_{11} = (q, q', q')$, $\bQ_{21} = (q, -q', -q')$, 
$\bQ_{31} = (-q, q', -q')$, $\bQ_{41} = (-q, -q', q')$, 
$\bQ_{12} = (q', q, q')$, $\bQ_{22} = (-q', q, -q')$, 
$\bQ_{32} = (-q', -q, q')$, $\bQ_{42} = (q', -q, -q')$, 
$\bQ_{13} = (q', q', q)$, $\bQ_{23} = (-q', -q', q)$, 
$\bQ_{33} = (q', -q', -q)$, and $\bQ_{43} = (-q', q', -q)$. 
The wave numbers $q$, $q^{\prime}$ are $q=0.1$ and $q^{\prime}=2q$ around $(n_{e}, U)=(1.75, 5)$.

The 12QMSDW are regarded as a superposition of three 4QMSDW's with $x$, $y$, $z$ polarizations, respectively.  In order to understand the magnetic structure of the 12QMSDW, we consider the first two 4QMSDW in Eq. (\ref{12qsdw}) (2-4QMSDW) in the following section.

\subsubsection{Magnetic structure of the 2-4QMSDW}

Let us consider first the following 2-4QMSDW.
\begin{align}
\bm_{l} &= 
m \boldsymbol{i} \big[ -\cos (\bQ_{11} \! \cdot \! \boldsymbol{R}_{l} + \delta) +
\cos (\bQ_{21} \! \cdot \! \boldsymbol{R}_{l}  + \delta) \nonumber \\
&\ \ \ \ \ \ \ \ \ +
\cos (\bQ_{31} \! \cdot \! \boldsymbol{R}_{l}  + \delta) - 
\cos (\bQ_{41} \! \cdot \! \boldsymbol{R}_{l}  + \delta) \big] \nonumber \\
& +
m \boldsymbol{j} \big[ -\cos (\bQ_{12} \! \cdot \! \boldsymbol{R}_{l}  + \delta) +
\cos (\bQ_{22} \! \cdot \! \boldsymbol{R}_{l}  + \delta) \nonumber \\
&\ \ \ \ \ \ \ \ \ +
\cos (\bQ_{32} \! \cdot \! \boldsymbol{R}_{l}  + \delta) - 
\cos (\bQ_{42} \! \cdot \! \boldsymbol{R}_{l}  + \delta) \big] \ .
\label{t4qsdw}
\end{align}
We note that the 2-4QMSDW is written as follows by using the sum-to-product formula.
\begin{align}
\bm_{l} &= 
2m \boldsymbol{i} \big[ \sin (qx_{l} + \delta) \sin q^{\prime} (y_{l}  + z_{l}) +
\sin (qx_{l}  - \delta) \sin q^{\prime} (y_{l}  - z_{l}) \big]    \nonumber \\
&+
2m \boldsymbol{j} \big[ \sin (qy_{l} + \delta) \sin q^{\prime} (x_{l}  + z_{l}) +
\sin (qy_{l}  - \delta) \sin q^{\prime} (-x_{l}  + z_{l}) \big] \ .
\label{t4qsdw3}
\end{align}

The 2-4QMSDW contains vortex structures.  In order to clarify this feature, we assume that the MSDW forms the planar vortex structures.  The conditions for their core centers $\boldsymbol{R}_{\rm c} = (x_{0}, y_{0}, z_{0})$ are given by
\begin{align}
\sin \Big( qx_{0} - \frac{\pi}{4} \Big) \sin q^{\prime} (z_{0} + y_{0}) -
\cos \Big(qx_{0} - \frac{\pi}{4} \Big) \sin q^{\prime} (z_{0} - y_{0}) = 0 \, ,  
\label{t4qccond1} \\
\sin \Big(qy_{0} - \frac{\pi}{4} \Big) \sin q^{\prime} (x_{0}  + z_{0}) +
\cos \Big(qy_{0} - \frac{\pi}{4} \Big) \sin q^{\prime} (z_{0} - x_{0}) = 0 \ .
\label{t4qccond2}
\end{align}
For $z_{0}$ such that $q^{\prime}z_{0} = n\pi$, the above equations reduce to
\begin{align}
\Big[ \sin \Big( qx_{0} - \frac{\pi}{4} \Big) +
\cos \Big( qx_{0} - \frac{\pi}{4} \Big) \Big] \sin q^{\prime} y_{0} = 0 \, ,  
\label{t4qccond3} \\
\Big[ \sin \Big( qy_{0} - \frac{\pi}{4} \Big) -
\cos \Big( qy_{0} - \frac{\pi}{4} \Big) \Big] \sin q^{\prime} x_{0} = 0 \ .
\label{t4qccond4}
\end{align}

We find 4 possible sets of solutions for Eqs. (\ref{t4qccond3}) and (\ref{t4qccond4});
$\boldsymbol{R}_{\rm c}  =  (l_{1}a/2q, \, (l_{2}+1/2)a/2q, na/2q^{\prime})$, 
$(l_{1}a/2q, y_{0}, na/2q^{\prime})$, 
$(x_{0}, l_{2}a/2q^{\prime}, na/2q^{\prime})$, and 
$(l_{1}a/2q^{\prime}, l_{2}a/2q^{\prime}, na/2q^{\prime})$, where $l_{1}$, $l_{2}$, and $n$ are integer.
Among them, the second and third solutions are not suitable as the core centers because they contain arbitrary parameters, $x_{0}$ and $y_{0}$.  In the first case, we do not find the vortex structure around $\boldsymbol{R}_{\rm c}$.  In what follows, we consider the last case:
\begin{align}
\boldsymbol{R}_{\rm c}  =  
\big( l_{1}, \, l_{2}, \, n \big) \frac{a}{2q^{\prime}}\ .
\label{t4qrc}
\end{align}

As in Sect. 3.1, we introduce the position vector $\boldsymbol{r}_{l}$ such that $\boldsymbol{R}_{l}=\boldsymbol{r}_{l}+\boldsymbol{R}_{\rm c}$, and expand Eq. (\ref{t4qsdw}) with respect to $\boldsymbol{Q} \cdot \boldsymbol{r}_{l}$.
When $q^{\prime}=kq$, we obtain $\bm_{l}$ in the lowest order as follows.
\begin{align}
\bm_{l} &= 
(-)^{l_{2}+n} \sqrt{2} m \boldsymbol{i} \, ( -c^{(-)}_{l_{1}}(k) \boldsymbol{Q}_{4} + 
c^{(+)}_{l_{1}}(k) \boldsymbol{Q}_{5} ) \cdot \boldsymbol{r}_{l}    \nonumber  \\
& \ + (-)^{l_{1}+n} \sqrt{2} m \boldsymbol{j} \, ( -c^{(-)}_{l_{2}}(k) \boldsymbol{Q}_{6} + 
c^{(+)}_{l_{2}}(k) \boldsymbol{Q}_{7} ) \cdot \boldsymbol{r}_{l} \ .
\label{t4qsdw4}
\end{align}
Here $c^{(\pm)}_{l}(k) = \cos l\pi/k \pm \sin l\pi/k$, $\boldsymbol{Q}_{4}=(0, q^{\prime}, q^{\prime})$, $\boldsymbol{Q}_{5}=(0, q^{\prime}, -q^{\prime})$, $\boldsymbol{Q}_{6}=(q^{\prime}, 0, q^{\prime})$, and $\boldsymbol{Q}_{7}=(-q^{\prime}, 0, q^{\prime})$.

When $k=2$ as in the case of $(n_{e}, U)=(1.75, 5)$, we have $\boldsymbol{R}_{\rm c}$ as 
\begin{align}
\boldsymbol{R}_{l_{1} l_{2}n}  =  
\Big( \frac{l_{1}}{4}, \, \frac{l_{2}}{4}, \, \frac{n}{4} \Big) \, L \ .
\label{t4qrc2}
\end{align}
Here $L=a/q$ denotes the longest wave length of the system.
We then examine the local magnetic structures $\bm_{l}$ around 16 independent $\boldsymbol{R}_{l_{1} l_{2}n}$ on a (001) plane with $n$.
We obtained around $\boldsymbol{R}_{00n}$, $\boldsymbol{R}_{02n}$, $\boldsymbol{R}_{20n}$,  and $\boldsymbol{R}_{22n}$,  the shear-flow type SDW structures with a node on a line through $\boldsymbol{R}_{\rm c}$ :
\begin{align}
\bm_{l} &= 
(-)^{n+1} 4\sqrt{2} mq (\pm\hat{z}_{l} \boldsymbol{i} \pm\hat{x}_{l} \boldsymbol{j} ) \ .
\label{t4qshear1}
\end{align}
Here the signs should be taken as $(+,+)$ for $\boldsymbol{R}_{00n}$, $(+,-)$ for $\boldsymbol{R}_{02n}$, $(-,+)$ for $\boldsymbol{R}_{20n}$,  and  $(-,-)$ for $\boldsymbol{R}_{22n}$.
Around $\boldsymbol{R}_{11n}$, $\boldsymbol{R}_{13n}$, $\boldsymbol{R}_{31n}$, and $\boldsymbol{R}_{33n}$, we have
\begin{align}
\bm_{l} &= 
(-)^{n+1} 4\sqrt{2} mq (\pm\hat{y}_{l} \boldsymbol{i} \pm\hat{z}_{l} \boldsymbol{j} ) \ .
\label{t4qshear2}
\end{align}
Here the signs should be taken as $(+,+)$ for $\boldsymbol{R}_{11n}$, $(+,-)$ for $\boldsymbol{R}_{13n}$, $(-,+)$ for $\boldsymbol{R}_{31n}$,  and  $(-,-)$ for $\boldsymbol{R}_{33n}$.

We obtained the SDW local structures with a node on the (001) plane around $\boldsymbol{R}_{01n}$, $\boldsymbol{R}_{03n}$, $\boldsymbol{R}_{21n}$, and $\boldsymbol{R}_{23n}$.
\begin{align}
\bm_{l} &= 
(-)^{n} 4\sqrt{2} mq \hat{z}_{l} (\pm\boldsymbol{i} \pm\boldsymbol{j}) \ .
\label{t4qsdw0}
\end{align}
Here the signs should be taken as $(+,+)$ for $\boldsymbol{R}_{01n}$, $(+,-)$ for $\boldsymbol{R}_{03n}$, $(-,+)$ for $\boldsymbol{R}_{21n}$,  and $(-,-)$ for $\boldsymbol{R}_{23n}$.
These local structures around 12 points are not vortex structures.

On the other hand, we obtained the clockwise planar vortex structure around $\boldsymbol{R}_{12n}$:
\begin{align}
\bm_{l} &= (-)^{n} 4\sqrt{2}mq r_{l} \Big[ \cos \Big( \phi_{l} - \frac{\pi}{2} \Big) \boldsymbol{i} + \sin \Big( \phi_{l} - \frac{\pi}{2} \Big) \boldsymbol{j} \Big] \, ,
\label{2-4qvor1}
\end{align}
the anticlockwise planar vortex structure around $\boldsymbol{R}_{30n}$:
\begin{align}
\bm_{l} &= (-)^{n} 4\sqrt{2}mq r_{l} \Big[ \cos \Big( \phi_{l} + \frac{\pi}{2} \Big) \boldsymbol{i} + \sin \Big( \phi_{l} + \frac{\pi}{2} \Big) \boldsymbol{j} \Big] \, ,
\label{2-4qvor2}
\end{align}
and the antivortices around $\boldsymbol{R}_{10n}$ and $\boldsymbol{R}_{32n}$:
\begin{align}
\bm_{l} &= (-)^{n} 4\sqrt{2}mq r_{l} \Big[ \cos \Big( -\phi_{l} \pm \frac{\pi}{2} \Big) \boldsymbol{i} + \sin \Big( -\phi_{l} \pm \frac{\pi}{2} \Big) \boldsymbol{j} \Big] \, ,
\label{2-4qavor}
\end{align}
respectively.

Moreover, we can verify that $\bm_{l}=\boldsymbol{0}$ on the lines $(l_{1}L/2, y_{l}, nL/4)$ and $(x_{l}, (2l_{2}+1)L/4, nL/4)$.  The planar vortices and antivortices in the 2-4QMSDW are separated each other by these node lines with no magnetic moment.
The local structures of the 2-4QMSDW around 16 points on the $xy$ plane are summalized in Fig. 5.  We find the vortices around $\boldsymbol{R}_{120}$, $\boldsymbol{R}_{300}$, and $\boldsymbol{R}_{340}$, and the antivortices around $\boldsymbol{R}_{100}$, $\boldsymbol{R}_{140}$, and $\boldsymbol{R}_{320}$.  They are separated by the node lines mentioned above.
%
%
\begin{figure}[htbp]
\begin{center}
\includegraphics[width=7.5cm]{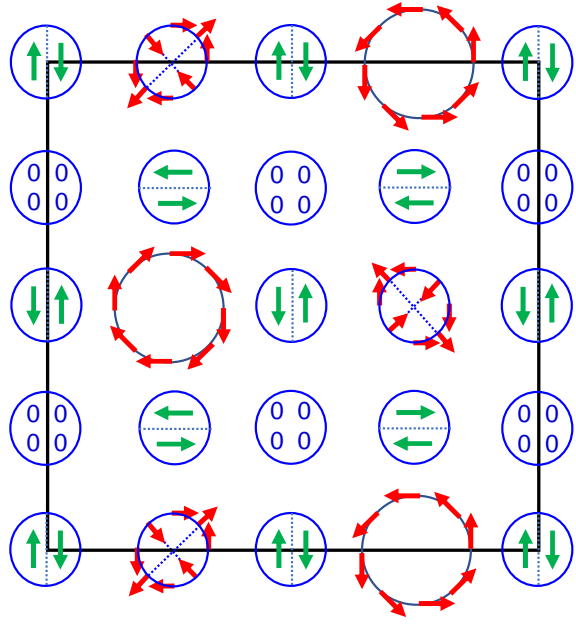}
\end{center}
\vspace{0cm}
\caption{ (Color online) Schematic diagram on the $L \times L$ $xy$ plane showing the local magnetic structures of the 2-4$Q$ multiple SDW (2-4QMSDW) with the wave numbers $q=0.1$ and  $q^{\prime}=2q$.  Here $L=a/q=10a$, $a$ being the lattice constant.  The same symbols as in Fig.2 are used for the clockwise vortex structure centered at $\boldsymbol{R}_{120}=(1/4, 1/2, 0)L$, the anticlockwise vortex centered at $\boldsymbol{R}_{300}=(3/4, 0, 0)L$, the antivortex centered at $\boldsymbol{R}_{100}=(1/4, 0, 0)L$, and the antivortex centered at $\boldsymbol{R}_{320}=(3/4, 1/2, 0)L$. The shear-flow type SDW structures with the node on the line parallel to the $y$ axis are shown by the symbol with the antiparallel arrows and vertical dotted line.  They are located at $\boldsymbol{R}_{000}$, $\boldsymbol{R}_{020}$, $\boldsymbol{R}_{200}$, and $\boldsymbol{R}_{220}$.  The shear-flow type SDW structures with the node on the line parallel to the $x$ axis, which are located at $\boldsymbol{R}_{110}$, $\boldsymbol{R}_{130}$, $\boldsymbol{R}_{310}$, and $\boldsymbol{R}_{330}$, are shown by the symbol with the antiparallel arrows and horizontal dotted line.  The SDW local structures with a node on the (001) plane are located at $\boldsymbol{R}_{010}$, $\boldsymbol{R}_{030}$, $\boldsymbol{R}_{210}$, and $\boldsymbol{R}_{230}$, and are shown by the symbol with four zeros (0) in a circle.
}
\label{fg524qmsdw}
\end{figure}
%
%

\subsubsection{Magnetic structure of 12$Q$MSDW}

We can analyze the 12QMSDW given by Eq. (\ref{12qsdw}) as a superposition of the 2-4QMSDW on the $xy$ plane (the first two terms in Eq. (\ref{12qsdw})) and the remaining 1-4QMSDW in the $z$ direction (the last term in Eq. (\ref{12qsdw})).
For example, around $\boldsymbol{R}_{12n}=(1/4, 1/2, n/4)L$, we have a planar vortex structure of 2-4QMSDW given by Eq. (\ref{2-4qvor1}) for $q^{\prime}=2q$.  The magnetic moments of the remaining 1-4QMSDW, $\bm_{l}$(1-4Q), can be expressed after transformation $\boldsymbol{R}_{l}=\boldsymbol{r}_{l}+\boldsymbol{R}_{12n}$ as
\begin{align}
\bm_{l}(\text{1-4Q}) &= 
-m \boldsymbol{k} \Big[ -\cos \Big( \bQ_{13} \! \cdot \! \boldsymbol{r}_{l}  + \frac{n\pi}{2} - \frac{\pi}{4} \Big) +
\cos \Big( \bQ_{23} \! \cdot \! \boldsymbol{r}_{l}  + \frac{n\pi}{2} - \frac{\pi}{4} \Big) \nonumber \\
&\ \ \ \ \ \ \ \ \ \ \ \ +
\cos \Big( \bQ_{33} \! \cdot \! \boldsymbol{r}_{l}  - \frac{n\pi}{2} - \frac{\pi}{4} \Big) - 
\cos \Big( \bQ_{43} \! \cdot \! \boldsymbol{r}_{l}  - \frac{n\pi}{2} - \frac{\pi}{4} \Big) \Big] \ .
\label{1-2qsdw}
\end{align}
Expanding $\bm_{l}$(1-4Q) with respect to $\bQ \cdot \boldsymbol{r}_{l}$, we obtain
\begin{align}
\bm_{l}(\text{1-4Q}) = (-)^{\frac{n}{2}} 2\sqrt{2} m q^{\prime} \hat{y}_{l} \, \boldsymbol{k} \ ,
\label{1-2qsdw2}
\end{align}
for even $n$, and 
\begin{align}
\bm_{l}(\text{1-4Q}) = (-)^{\frac{n+1}{2}} 2\sqrt{2} m q^{\prime} \hat{x}_{l} \, \boldsymbol{k} \ ,
\label{1-2qsdw3}
\end{align}
for odd $n$.

Superposing Eq. (\ref{1-2qsdw2}) or Eq. (\ref{1-2qsdw3}) on Eq. (\ref{2-4qvor1}), we obtain the 12QMSDW for $q^{\prime}=2q$ around $\boldsymbol{R}_{12n}$ as follows.
\begin{align}
\bm_{l}^{(12n)} &= (-)^{n} 4\sqrt{2}mq r_{l} \Big[ \cos \Big( \phi_{l} - \frac{\pi}{2} \Big) \boldsymbol{i} + \sin \Big( \phi_{l} - \frac{\pi}{2} \Big) \boldsymbol{j} \Big] \nonumber \\
&+ \begin{cases}
(-)^{\frac{n}{2}} 4\sqrt{2} m q \hat{y}_{l} \, \boldsymbol{k} & \text{($n=$even)} \ ,
\vspace{1mm} \\
(-)^{\frac{n+1}{2}} 4\sqrt{2} m q \hat{x}_{l} \, \boldsymbol{k} & \text{($n=$odd)} \ .
\end{cases}
\label{12qvor12}
\end{align}
This is a vortex structure in which the $z$ components of the magnetic moments are polarized in opposite directions across the line $y=L/2$ or $x=L/4$ through $\boldsymbol{R}_{12n}=(1/4,1/2,n/4)L$.  Consequently, there is no core polarization in the vortex, and thus it is not the skyrmion.

Similarly, we obtain another vortex structure with opposite helicity around $\boldsymbol{R}_{30n}=(3/4, 0, n/4)L$ as
\begin{align}
\bm_{l}^{(30n)} &= (-)^{n} 4\sqrt{2}mq r_{l} \Big[ \cos \Big( \phi_{l} + \frac{\pi}{2} \Big) \boldsymbol{i} + \sin \Big( \phi_{l} + \frac{\pi}{2} \Big) \boldsymbol{j} \Big] \nonumber \\
&+ \begin{cases}
(-)^{\frac{n}{2}} 4\sqrt{2} m q \hat{y}_{l} \, \boldsymbol{k} & \text{($n=$even)} \ ,
\vspace{1mm} \\
(-)^{\frac{n+1}{2}} 4\sqrt{2} m q \hat{x}_{l} \, \boldsymbol{k} & \text{($n=$odd)} \ .
\end{cases}
\label{12qvor30}
\end{align}

The antivortex structures are obtained around $\boldsymbol{R}_{32n}=(3/4, 1/2, n/4)L$ and $\boldsymbol{R}_{10n}=(1/4, 0, n/4)L$ as follows.
\begin{align}
\bm_{l}^{(32n)} &= (-)^{n} 4\sqrt{2}mq r_{l} \Big[ \cos \Big( -\phi_{l} - \frac{\pi}{2} \Big) \boldsymbol{i} + \sin \Big( -\phi_{l} - \frac{\pi}{2} \Big) \boldsymbol{j} \Big] \nonumber \\
&+ \begin{cases}
(-)^{\frac{n}{2}} 4\sqrt{2} m q \hat{y}_{l} \, \boldsymbol{k} & \text{($n=$even)} \ ,
\vspace{1mm} \\
(-)^{\frac{n+1}{2}} 4\sqrt{2} m q \hat{x}_{l} \, \boldsymbol{k} & \text{($n=$odd)} \ ,
\end{cases}
\label{12qvor32}
\end{align}
\begin{align}
\bm_{l}^{(10n)} &= (-)^{n} 4\sqrt{2}mq r_{l} \Big[ \cos \Big( -\phi_{l} + \frac{\pi}{2} \Big) \boldsymbol{i} + \sin \Big( -\phi_{l} + \frac{\pi}{2} \Big) \boldsymbol{j} \Big] \nonumber \\
&+ \begin{cases}
(-)^{\frac{n}{2}} 4\sqrt{2} m q \hat{y}_{l} \, \boldsymbol{k} & \text{($n=$even)} \ ,
\vspace{1mm} \\
(-)^{\frac{n+1}{2}} 4\sqrt{2} m q \hat{x}_{l} \, \boldsymbol{k} & \text{($n=$odd)} \ .
\end{cases}
\label{12qvor10}
\end{align}
These antivortices also have no core polarization at $\boldsymbol{R}_{32n}$ and $\boldsymbol{R}_{10n}$. 

It should be noted that Eq. (\ref{12qsdw}) for the 12QMSDW has the cyclic symmetry for $x$, $y$, $z$ variables.  Thus the same types of vortex structures appear on the $yz$ and $zx$ planes.  The expressions of these structures on the $yz$ plane are obtained by replacement $(\hat{x}_{l}, \hat{y}_{l}, \hat{z}_{l}) \rightarrow (\hat{y}_{l}, \hat{z}_{l}, \hat{x}_{l})$ and $(\boldsymbol{i}, \boldsymbol{j}, \boldsymbol{k}) \rightarrow (\boldsymbol{j}, \boldsymbol{k}, \boldsymbol{i})$ in Eqs. (\ref{12qvor12}) $\sim$ (\ref{12qvor10}).  Those on the $zx$ plane are obtained by further replacement  $(\hat{y}_{l}, \hat{z}_{l}, \hat{x}_{l}) \rightarrow (\hat{z}_{l}, \hat{x}_{l}, \hat{y}_{l})$ and $(\boldsymbol{j}, \boldsymbol{k}, \boldsymbol{i}) \rightarrow (\boldsymbol{k}, \boldsymbol{i}, \boldsymbol{j})$.
%
%
\begin{figure}[htbp]
\begin{center}
\includegraphics[width=11cm]{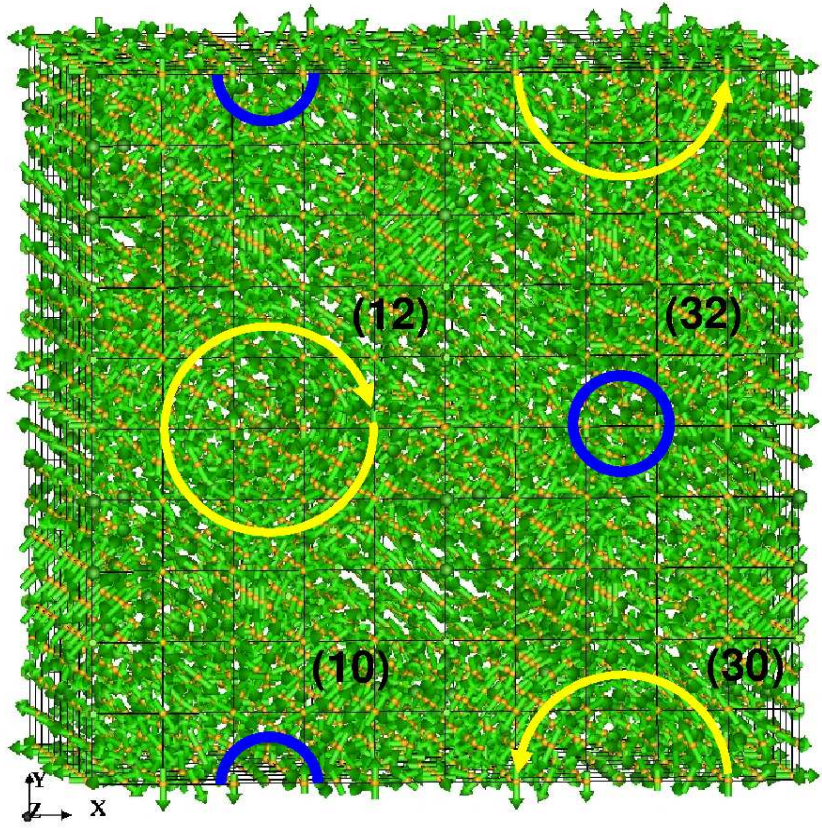}
\end{center}
\vspace{0cm}
\caption{ (Color online)  12$Q$ multiple SDW (12QMSDW) obtained by the GHF calculations for $(n_{e},U) = (1.75, 5)$.  The clockwise (anticlockwise) vortex structure centered at $\boldsymbol{R}_{120}$ ($\boldsymbol{R}_{300}$) on the $xy$ plane is shown by the yellow arrow circle, while the antivortex structures centered at $\boldsymbol{R}_{320}$ and $\boldsymbol{R}_{100}$ are shown by the blue circles.  The index $(lm)$ for each vortex expresses its core center position $\boldsymbol{R}_{lm0}$.  See Fig. \ref{fg712qmsdw} for comparison with the analytic result.  Use a zoom-in tool to see more detailed structure in the AVS image.
}
\label{fg612qmsdw}
\end{figure}
%
%
\begin{figure}[htbp]
\begin{center}
\includegraphics[width=17cm]{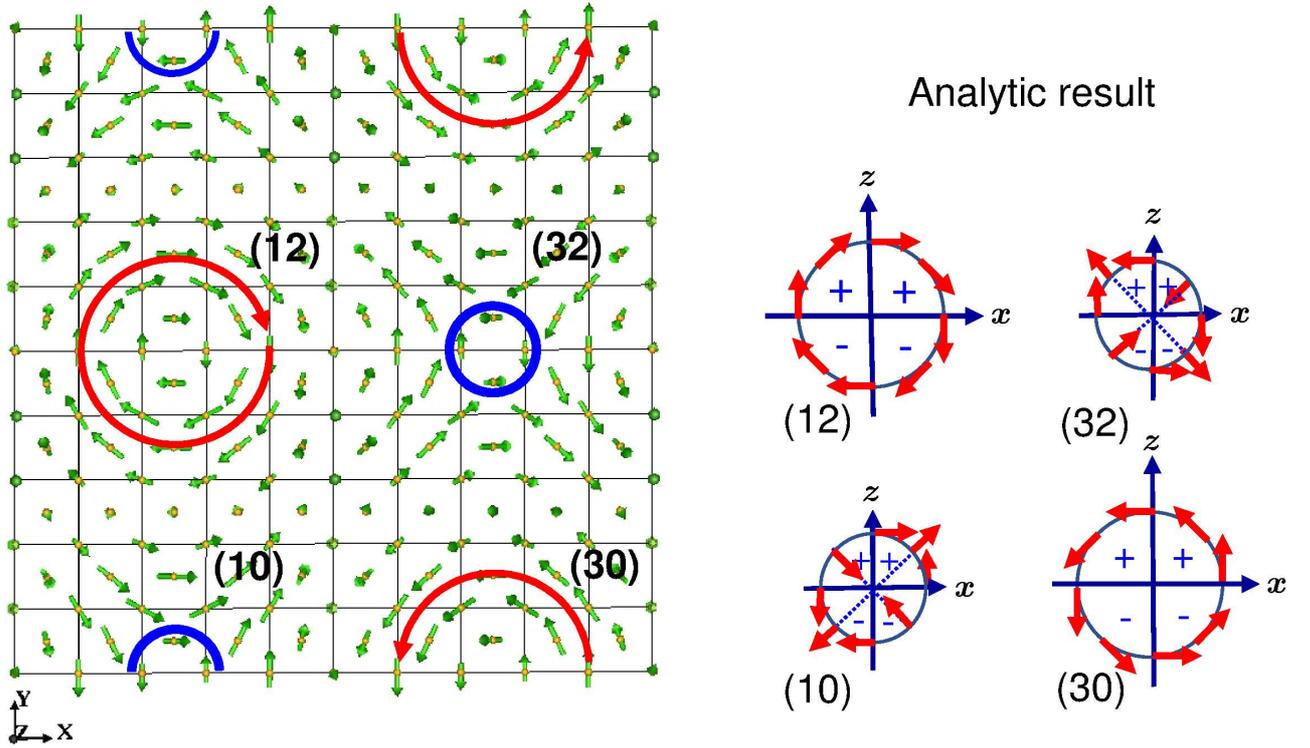}
\end{center}
\vspace{0cm}
\caption{ (Color online)  Magnetic structure of the 12QMSDW on the $z_{l}=0$ plane for $(n_{e},U) = (1.75, 5)$ (left) and its analytic result (right).  At the left-hand-side, the clockwise (anticlockwise) vortex structure centered at $\boldsymbol{R}_{120}$ ($\boldsymbol{R}_{300}$) is shown by the red arrow circle, while the antivortex structures centered at $\boldsymbol{R}_{320}$ and $\boldsymbol{R}_{100}$ are shown by the blue circles.  The same symbols as in Fig.2 are used in the analytic result at the right-hand-side.  The index $(lm)$ for each vortex expresses its core center position $\boldsymbol{R}_{lm0}$.   Use a zoom-in tool to see more detailed structure in the AVS image.
}
\label{fg712qmsdw}
\end{figure}
%
%

The vortex structure around $\boldsymbol{R}_{n12}=(n/4, 1/4, 1/2)L$ on the $yz$ plane, for example, is expressed as
\begin{align}
\bm_{l}^{(n12)} &= (-)^{n} 4\sqrt{2}mq r_{l} \Big[ \cos \Big( \phi_{l} - \frac{\pi}{2} \Big) \boldsymbol{j} + \sin \Big( \phi_{l} - \frac{\pi}{2} \Big) \boldsymbol{k} \Big] \nonumber \\
&+ \begin{cases}
(-)^{\frac{n}{2}} 4\sqrt{2} m q \hat{z}_{l} \, \boldsymbol{i} & \text{($n=$even)} \ ,
\vspace{1mm} \\
(-)^{\frac{n+1}{2}} 4\sqrt{2} m q \hat{y}_{l} \, \boldsymbol{i} & \text{($n=$odd)} \ ,
\end{cases}
\label{12qvorn12}
\end{align}
and the vortex structure around $\boldsymbol{R}_{2n1}=(1/2, n/4, 1/4)L$ on the $zx$ plane is given by
\begin{align}
\bm_{l}^{(2n1)} &= (-)^{n} 4\sqrt{2}mq r_{l} \Big[ \cos \Big( \phi_{l} - \frac{\pi}{2} \Big) \boldsymbol{k} + \sin \Big( \phi_{l} - \frac{\pi}{2} \Big) \boldsymbol{i}) \Big] \nonumber \\
&+ \begin{cases}
(-)^{\frac{n}{2}} 4\sqrt{2} m q \hat{x}_{l} \, \boldsymbol{j} & \text{($n=$even)} \ ,
\vspace{1mm} \\
(-)^{\frac{n+1}{2}} 4\sqrt{2} m q \hat{z}_{l} \, \boldsymbol{j} & \text{($n=$odd)} \ .
\end{cases}
\label{12qvor2n1}
\end{align}
Here the polar coordinates $(r_{l}, \phi_{l})$ are defined on the $yz$ ($zx$) plane in Eq. (\ref{12qvorn12}) (Eq. (\ref{12qvor2n1})).

In Figs. 6 and 7, we compare the AVS image with the analytic result for the 12QMSDW $(n=1.75, U=5)$ on the $xy$ plane ($n=0$).  We find that the present analysis explains well the existence of the clockwise and anticlockwise vortex structures around $\boldsymbol{R}_{120}$ and $\boldsymbol{R}_{300}$, respectively.  The magnetic moments at the upper half of these vortices are polarized in the $z$ direction, while those at the lower half are polarized in the $-z$ direction.  In particular, there is no polarization at their core centers.  Thus these vortices are not half skyrmions, though we suggested the latter in paper I.
We also verify in the AVS image the antivortices around $\boldsymbol{R}_{320}$ and $\boldsymbol{R}_{100}$, as found in the present analysis.  The vortices on the $xy$ plane changes their sign alternatively with increasing the layer number $n$, and the positive (negative) regions of the $z$ polarization in these vortices rotate anticlockwise with increasing $n$.

\section{Summary}

We have examined the analytic property of the multiple spin density waves (MSDW) showing the vortex structures, which were found in the AVS image analysis for the magnetic structures of the Hubbard model on the fcc lattice obtained by the generalized Hartree-Fock  (GHF) approximation~\cite{kake20}, in order to understand the vortex structures and their relations to the MSDW in the itinerant electron system with inversion symmetry.  

The vortex structures in the MSDW are accompanied by the vortex cores $\boldsymbol{R}_{\rm c}$.  We determined $\boldsymbol{R}_{\rm c}$ solving their equations.  Expanding the magnetic moments $\boldsymbol{m}_{l}$ for the MSDW with respect to the position vector around the core centers  $\boldsymbol{R}_{\rm c}$, we obtained the analytic expressions for the vortex structures.

In the 2$Q$ multiple helical SDW (2QH) with wave number $q=0.2$, which was found at $(n_{e}, U)=(1.7, 5.5)$ in the GHF calculations, we determined the two types of vortex structures with different helicity ($\pm\pi/2$) and core polarization ($\pm2m$) on the $xy$ plane.  These are the half-skyrmion and half-antiskyrmion particles, forming the antiferromagnetic (AF) structure on a  giant face-centered square lattice.
We also found the antivortex structures with zero core polarization between the half skyrmions with the same helicity, which were overlooked in our previous AVS image analyses.

The half-skyrmion lattice structure has recently been reported in the Co${}_{8}$Zn${}_{9}$Mn${}_{3}$ compound~\cite{yu18}.  The AF half-skyrmion vortex lattice structure agrees with that was observed in Co${}_{8}$Zn${}_{9}$Mn${}_{3}$.  However the antivortices with no core polarization between the half-skyrmions have not yet been verified experimentally.  In the experimental analysis, the half-antiskyrmions with positive core polarization are allocated from the viewpoint of the local-moment model~\cite{yu18}.  Experimental data are obtained under the external field 20mT being perpendicular to the $(001)$ plane.  Detailed experimental analyses under zero magnetic field are desired to resolve the discrepancy, though it is not easy to determine the out-of-plane magnetization by means of the Lorenz transmission electron microscopy technique. 

The 3$Q$ multiple helical SDW (3QH) found at $(n_{e}, U)=(1.68, 6)$, $(1.6, 6.5)$, $(1.42, 7.5)$ is a superposition of the 2QH and the 1QH ``westerlies'' on the (001) plane.  On the basis of the 2QH analysis, we verified that the 2QH vortex and antivortex structures with ``westerlies'' on the $(001)$ plane are twisted along the $[001]$ direction in the 3QH.
The 3QH structure is suggested to be realized in the MnGe compound~\cite{zhan16}.  Detailed experimental analyses of the magnetic structure to verify the local vortex structures on the (001) plane, however, have not yet been reported.

The 12$Q$ multiple SDW (12QMSDW) stabilized around $(n_{e}, U)=(1.75, 5)$ are regarded as the superposition of the 2-4QMSDW and the remaining 4QMSDW.  We found analytically that the 2-4QMSDW forms a large vortex lattice structure on the $(001)$ plane, in which the clockwise vortices with no core polarization occupy the corners and the anticlockwise vortices with no core polarization occupy the centers on the face-centered square lattice with lattice constant $L(=a/q)$.  Consequently, we showed that the same types of vortex lattice structures appear on the $xy$ plane even in the 12QMSDW, though these vortices are accompanied by additional positive (negative) polarization in the $z$ direction in the upper (lower) half plane of each vortex.  The vortices in the 12QMSDW, therefore, are not the half skyrmions, though we suggested the latter in paper I.  The antivortices appear between the vortices with the same helicity.  

We also clarified that the vortex lattice structure on the $xy$ plane appears every $5a/2$ translation along $[001]$ direction in the 12QMSDW, and the signs of the magnetic moments on the $xy$ plane change when the vortex layer moves up or down by one.  Moreover we pointed out that the same vortex structures appear on the $yz$ and $zx$ planes due to  cyclic symmetry of the 12QMSDW.
The 12QMSDW showing vortex structures have not yet been found experimentally.
Experimental search for the structure is left for future work.

The method presented here provides us with a simple tool to understand the complex magnetic structures of the nano-meter scale from a microscopic point of view.  Applications of the present approach combined with the visualization image techniques to the other itinerant electron systems are desirable for future investigations of the nano-scale magnetic structure and magnetism.

\begin{acknowledgment}


The author would like to express his sincere thanks to Prof. T. Uchida
 for valuable discussions and comments on the present work. 

\end{acknowledgment}



\end{document}